\documentclass[a4paper,11pt]{article}
\usepackage{amsfonts,latexsym}
\usepackage{graphicx}
\usepackage{dcolumn}
\usepackage{bm}
\newcommand{\be}{\begin{equation}}
\newcommand{\ee}{\end{equation}}
\newcommand{\ba}{\begin{eqnarray}}
\newcommand{\ea}{\end{eqnarray}}

\title{Towards a precise determination of the topological \\
       susceptibility in the SU(3) Yang-Mills theory }

\author{Leonardo Giusti\\
        CERN, Physics Department, 1211 Geneva 23, Switzerland \\
        Dipartimento di Fisica, Universit\`a di Milano Bicocca,  \\
        Piazza della Scienza 3, I-20126 Milano, Italy\\
        E-mail: {leonardo.giusti@cern.ch}
        \and
   Bruno~Taglienti\\
   INFN, Sezione di Roma, P.le A. Moro 2, I-00185 Roma, Italy\\
   E-mail: {Bruno.Taglienti@roma1.infn.it}
   \and
   Silvano~Petrarca \\
   Dip. di Fisica, Universit\`a di Roma ``La Sapienza'',\\ P.le A. Moro 2, I-00185 Rome, Italy\\
   INFN, Sezione di Roma, P.le A. Moro 2, I-00185 Roma, Italy\\	 
   E-mail: {Silvano.Petrarca@roma1.infn.it} }

\begin{document}
\maketitle

\begin{abstract}
An ongoing effort to compute the topological susceptibility 
for the $SU(3)$ Yang-Mills theory in the continuum limit with a 
precison of about $2\%$ is reported. The susceptibility is 
computed by using the definition of the charge suggested by 
Neuberger fermions for two values of the negative mass parameter s. 
Finite volume and discretization effects are estimated to meet this
level of precision. The large statistics required has been obtained by 
using PCs of the INFN-GRID.Simulations with 
larger lattice volumes are necessary in order to 
better understanding the continuum limit at small lattice spacing values.

\end{abstract}


\section{Introduction}
\noindent We continue our numerical study on the lattice of the topological charge 
distribution in the $SU(3)$ Yang-Mills theory adopting the definition suggested 
by Neuberger fermions as discussed in a series of papers~\cite{Neu1}-\cite{Lus2}.\\
\indent Recent numerical studies on this subject can be found in 
refs.\cite{Edwards:1998wx,GLWW03,DP} and ref.~\cite{DGP04}. In the last paper a 
systematic study at different volumes and values of the lattice spacing was performed 
in order to obtain a reliable determination of the topological susceptibility, 
the first cumulant of the distribution, at the $5\%$ level in the continuum limit
at infinite volume. The result supports the Witten-Veneziano explanation 
for the large mass of the ${\eta}'$.\\
\indent The aim of refs \cite{secondo,terzo} was to look for 
non-gaussianities in the topological charge distribution of the $SU(3)$ 
Yang-Mills theory. This is particularly challenging since the contribution
of the $n^\mathrm{th}$ cumulant to the charge distribution is suppressed as 
$V^{n-1}$ in its asymptotic expansion, even though the cumulant itself is a 
quantity of $O(1)$ in the infinite volume limit. It is also worth noting that 
in order to search for such very small sub-leading effects it is necessary to 
be sure that all the systematics of the calculation cannot either simulate 
or hide the effect, and therefore a solid theoretical framework, such as 
the one provided by the topological charge  definition suggested from 
Ginsparg-Wilson fermions, is indispensable. In ref.~\cite{DGP04} three 
different lattices at the same physical volume of $\sim (1.12 {\rm fm})^4$ 
but with about ten time more statistics than in ref.~\cite{DGP04}, i.e. 
roughly $3\cdot 10^4$ configurations, were studied in order to be able to unveil 
deviations from the Gaussian distribution. 
Significant deviations were found,  
the second cumulant of the distribution gets a value definitively 
different from zero within errors, and the distribution function agrees at that 
precision level with a first order modification in $V^{-1} $ of the Gaussian 
(Edgeworth expansion). The results clearly disfavour the $\theta$ behaviour of 
the vacuum energy predicted by dilute instanton models, while they are compatible 
with the expectation from the large $N_c$ expansion.\\

\noindent Here we present preliminary results for a measurement on the lattice of 
the topological susceptibility at the $2\%$ level. 
In order to keep finite volume and 
discretization effects below this level of precision,
we have exploited data produced in ref.~\cite{terzo} together with new data generated
 by additional lattices as described below.\\

\noindent  All the above challenging Monte Carlo calculation have been made possible 
by important improvements in algorithms which guarantee the reliability and the feasibility 
of high statistics. In particular we have used algorithms for zero mode counting with no 
contamination from quasi zero modes, optimized to run fast on a single processor~\cite{GHLW02}.
The considerable amount of computer time needed has been granted to us by the INFN GRID project. 
It allowed us to use the computer resources shared in the scientific Italian network provided 
by INFN along these years. We have also taken in advantage of the computer resources of the Italian 
organization COMETA.
 
 \section{Theoretical framework}
\noindent In this section we summarize our notation and the necessary theoretical framework, for a complete 
discussion see for example~\cite{terzo}. In the following we use the plaquette Wilson action of the 
$SU(3)$ gauge field. The massless lattice Neuberger-Dirac operator $D$ satisfies the Ginsparg-Wilson 
relation \cite{Neu1,Neu2} 
\begin{equation}
\gamma_5 D + D \gamma_5= \bar{a} D \gamma_5 D \; ,
\end{equation}
and the associated topological charge density can be defined as
\begin{equation}\label{eq:qx}
a^4 q(x) = -\frac{\bar a}{2}\, \mathrm{Tr}\Big[\gamma_5 D(x,x)\Big] ,
 \label{eq:chaden}
\end{equation}
where  ${\bar a}=a/(1+s)$,  $a$ is the lattice spacing and
$s$ is the negative mass parameter. The latter has been fixed in our calculation at the values 
$s=0.4$ for the study of the volume effects and to $s=0.0$ and $s=0.4$ for the study of the 
discretization behaviour. The topological charge is obtained from the lattice by computing on each
gauge configuration the number and the chirality of the zero modes of $D$ with the algorithm proposed 
in Ref.~\cite{GHLW02}. The index $\nu$ of the Dirac operator
\begin{equation}
\nu =n_+ - n_- 
\end{equation}
is directly related to the topological charge $Q$
 \begin{displaymath}
\nu=Q=a^4 \sum_x q(x)\; .
\end{displaymath}
 In the Euclidean space-time the ground-state energy $F(\theta)$ is
defined as
\be
e^{- F(\theta)} = \langle e^{i\theta Q}\rangle\; ,
\ee
where, as usual, $\langle\dots\rangle$ indicates the path-integral
average (our normalization is $F(0)=0$). In the large volume regime
$F(\theta)$ is proportional to the size $V$ of the system,
a direct consequence of the fact that the topological charge operator $Q$
is the four-dimensional integral of a local density.
The function $F(\theta)$ is related to the probability
of finding a gauge field configuration with
topological charge $Q=\nu$ by the Fourier transform
\be\label{eq:pnu}
P_\nu = \int_{-\pi}^\pi \frac{d\theta}{2 \pi}
e^{-i\theta\nu} e^{- F(\theta)}\; .
\ee
Large $N_c$ arguments
with $N_c$ being the number of colors, suggest that the fluctuations
of the topological charge are of quantum non-perturbative
nature. The $\theta$ dependence of the vacuum energy is expected
at leading order in $1/N_c$, and the normalized cumulants
\be\label{eq:cn}
{C}_{n} = (-1)^{n+1}\frac{1}{V}\frac{d^{2n}}{d\theta^{2n}}
F(\theta) \Big|_{\theta=0}\qquad n=1,2,\dots\; ,
\ee
which should scale asymptotically as $N_c^{2-2n}$, 
have to be determined with a non-perturbative 
computation. The normalized cumulants ${C}_{n}$ can thus be  
defined as the integrated connected correlation functions of $n$
charge densities (correlation functions
of an odd number of topological charges vanish thanks
to the invariance of the theory under parity):
\be\label{eq:CnQCd}
{C}_{n} = \frac{a^{8n}}{V} \sum_{x_1,\dots,x_{2n}}
\langle q(x_1)\dots q(x_{2n})\rangle^\mathrm{con}\; .
\ee
They have an unambiguous finite continuum limit which is
independent of the details of the regularization~\cite{GRTV,GRT,Lus2}.
At finite lattice spacing they are affected by discretization
errors which start at $O(a^2)$.

The Monte Carlo technique adopted here generates the gauge configurations
with a probability density proportional to $e^{-S_\mathrm{YM}}$, with
$S_\mathrm{YM}$ being the chosen discretization of the Yang--Mills action. This
algorithm performs an importance sampling of the topological charge with the
probability distribution given in Eq.~(\ref{eq:pnu}). A statistical signal for the
$n^\mathrm{th}$ cumulant is then obtained when the number of configurations
in the sample is
high enough  to be sensitive to terms suppressed
as $V^{n-1}$ in the asymptotic expansion. For instance, the estimators of the first two
cumulants
\ba
\overline{Q^2} & = & \frac{1}{N}
\sum_{i=1}^{N} \nu^2_i\; , \label{eq:est1}\\[0.125cm]
\overline{Q^{4,}} \,\!^\mathrm{con} & = &
\frac{1}{N}\sum_{i=1}^{N} \nu^4_i - 3 \left(\frac{1}{N}\sum_{i=1}^{N} \nu^2_i\right)^2\; ,
\label{eq:est2}
\ea
with $\nu_i$ being the value of the topological charge for a given gauge configuration
and $N$ the total number of configurations, have variances which, up to sub-leading
corrections, are given by $(2\sigma^4+\sigma^2\tau)/N$ and $(24\sigma^8 + 72 \sigma^6\tau)/N$
respectively being $\sigma^2 = V {C}_1$ and $\tau={C}_2/{C}_1$.

\section{Lattice data and results}
\begin{table}[!t]
\begin{center}
\begin{tabular}{llccccll}
\hline
Lat    &$\beta$&$L/a$&$r_0/a$&$L$[fm]&$N$& $\langle Q^2 \rangle$&$r_0^4 \chi$\\[0.125cm]
\hline
${\rm A}_1$&$6.0$   &$12$& $5.368$ &$1.12$&$34800$& $1.627(13)$ & $0.0652(12)$    \\[0.125cm]
${\rm B}_0$&$5.9138$&$12$& $4.601$ &$1.30$&$10000$& $3.271(47)$ & $0.0707(15)$\\[0.125cm]
${\rm B}_1$&$6.0$   &$14$& $5.368$ &$1.30$&$30000$& $3.097(26)$ & $0.0669(12)$ \\[0.125cm]
${\rm B}_2$&$6.0808$&$16$& $6.135$ &$1.30$&$10000$& $2.914(43)$ & $0.0630(14)$\\[0.125cm]
${\rm B}_3$&$6.1568$&$18$& $6.902$ &$1.30$&$10000$& $2.843(42)$ & $0.0615(14)$    \\[0.125cm]
${\rm C}_1$&$6.0$   &$16$& $5.368$ &$1.49$&$10000$& $5.314(75)$ & $0.0673(14)$    \\[0.125cm]
${\rm D}_1$&$6.0$   &$18$& $5.368$ &$1.60$&$10000$& $8.40(12) $ & $0.0664(14)$    \\[0.125cm]
\hline
${\cal B}_0$&$5.9138$&$12$& $4.601$ &$1.30$&$10000$& $2.501(36)$ & $0.0541(11) $\\[0.125cm]
${\cal B}_1$&$6.0$   &$14$& $5.368$ &$1.30$&$10000$& $2.761(40)$ & $0.0597(13)$\\[0.125cm]
${\cal B}_2$&$6.0808$&$16$& $6.135$ &$1.30$&$10000$& $2.723(40)$ & $0.0589(13)$\\[0.125cm]
${\cal B}_3$&$6.1568$&$18$& $6.902$ &$1.30$&$10000$& $2.788(40)$ & $0.0602(14)$\\
\hline
\end{tabular}
\caption{Simulation parameters and results.The first seven lattices are generated 
with the parameter value $s=0.4$, while the other four, under the horizontal
line, are for $s=0.0$. The topological susceptibility is given by $\chi={\langle Q^2\rangle / V}=C_1$.  \label{tab:newresults}}
\end{center}
\end{table}
\noindent The properties of the lattices considered and the results obtained 
for the topological susceptibility are reported in 
Table~\ref{tab:newresults}. 

\begin{figure}[h]
   \includegraphics[width=1\textwidth]{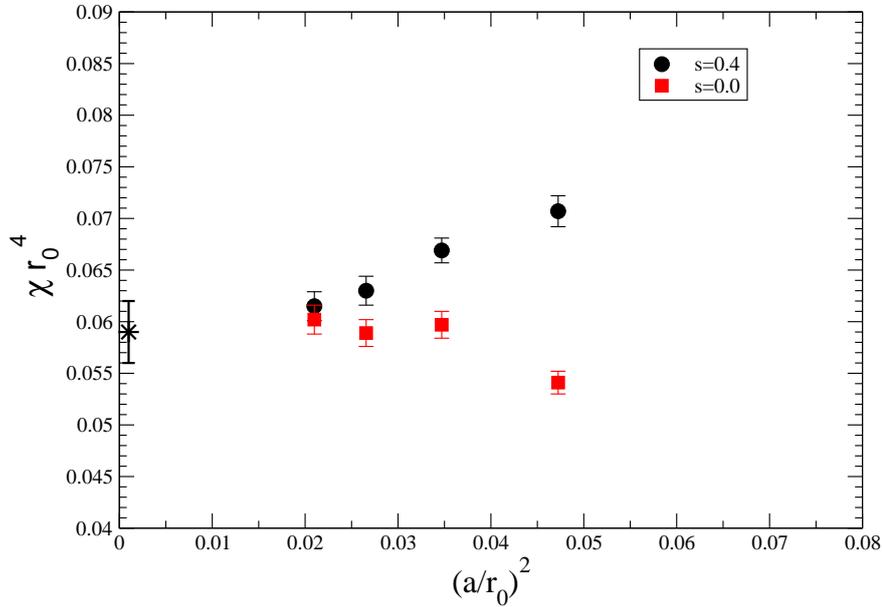}
   \caption{Topological susceptibility 
   as a function of the square of the lattice spacing at a fixed value 
            of the lattice size $L=1.304$~fm: data at mass parameter $s=0$ (squares) and $s=0.4$ (circles) 
            are shown. The continuum limit point is the result quoted in ref.~\cite{DGP04}.}
    \label{fig:discreto}
   \end{figure}

On the sets ${\rm A}_1, \, {\rm B}_0, \,  
{\rm B}_1, \, {\rm B}_2, \, {\rm B}_3, \, {\rm C}_1$ and ${\rm D}_1$ the topological charge 
is computed with $s=0.4$, while it is determined with $s=0.0$ on the lattices  
${\cal B}_0,\, {\cal B}_1,\, {\cal B}_2$ and ${\cal B}_3$.
  \begin{figure}[h]
   \includegraphics[width=1\textwidth]{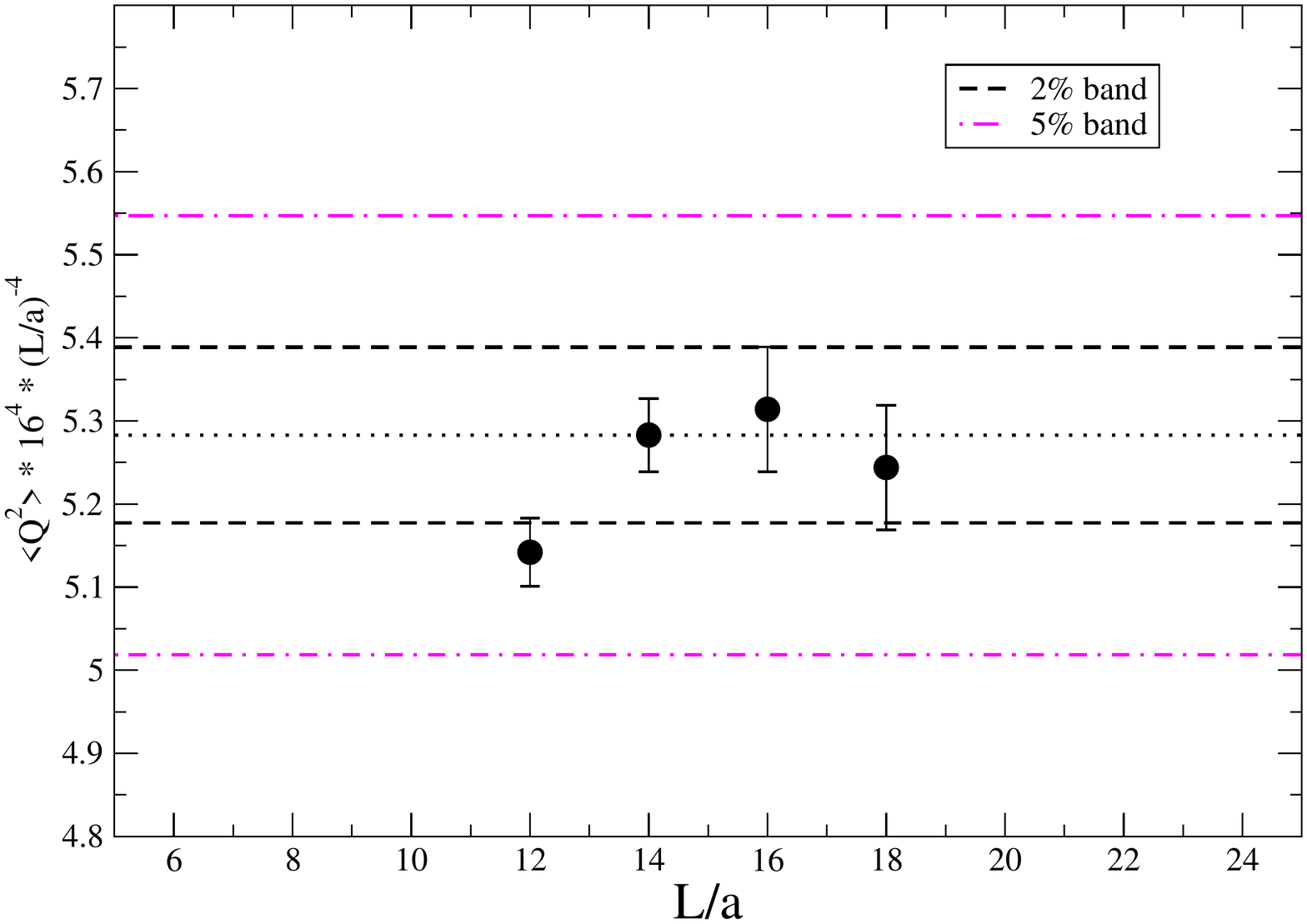}
   \caption{Rescaled topological charge as a function of the lattice size. Bands of $\pm 2\%$
    and $\pm 5\%$ centered at the value measured at $L=14$ are also shown.}
   \label{fig:volume}
   \end{figure}
In order to estimate the magnitude of finite size effects we have considered 
four lattices (${\rm A}_1$, ${\rm B}_1$, ${\rm C}_1$, ${\rm D}_1)$) at the same
value of the coupling constant corresponding to $\beta=6.0$ and at the same value 
of the negative mass parameter $s=0.4$. The values of the topological 
charge rescaled with respect to the reference volume of $16^4$ are shown
in Fig.~\ref{fig:volume}. It is rather clear that for these volumes 
the results are scattered in a $2\%$ band centered around the point with
$L/a=14$. While we cannot exclude that the point at $L/a=12$  turns out to be lower
with respect to the others due to a statistical fluctuation only, the other 
three points indicate clearly that finite size effects are within our statistical errors
for volumes larger or equal than $(1.3\; {\rm fm})^4$.\\
\indent Building on this result, the coupling constant of the other seven lattices
have been chosen so that the lattice linear extent is fixed to be $1.3$~fm, while the 
values of the coupling constant and of the negative mass $s$ are chosen in order to 
properly estimate discretization errors at this level of precision. The results 
for the topological susceptibility for all lattices with linear extent of $1.3$~fm
are shown in Fig.~\ref{fig:discreto} as a function of $(a/r_0)^2$. The data show a clear 
trend to converge to the same value in the continuum limit within the statistical precision 
reached. Nevertheless a better understanding of the behaviour at small lattice spacing values, 
i.e. larger lattice volumes, seems to be necessary in order to improve the continuum limit 
study. In fact, data at the $2\%$ level of error indicate a statistically non-negligible 
(and non-universal) contributions of terms of $O(a^4)$ to the behaviour of the two curves
in this range of values of $\beta$. Simulations at larger values of $\beta$ are underway. 
   
\section{Acknowledgments}
\noindent We warmly thank G. Andronico for the great organization of Theophys, the virtual organization 
of  INFN Grid Project for theoretical physics. We thank A. De Salvo and A. Colla of the INFN sez. Rome 
for the continuous effort in helping us during the accomplishment of the project and the COMETA 
organization for having opened their virtual organization to us.


\begin{thebibliography}{99}

\bibitem{Neu1} 
H.~Neuberger,
Phys. Lett.  {\bf B417}, 141, (1998), hep-lat/9707022.
\bibitem{Neu2} 
H.~Neuberger,
Phys. Rev.  {\bf D57}, 5417, (1998), hep-lat/9710089.
\bibitem{Has} 
P.~Hasenfratz, V.~Laliena, F.~Niedermayer,
Phys. Lett.  {\bf B427}, 125, (1998),  hep-lat/9801021.
\bibitem{Lus1}
M.~L\"uscher, 
Phys. Lett.  {\bf B428},  342, (1998),  hep-lat/9802011.
\bibitem{GRT}   
L.~Giusti, G.~C.~Rossi, M.~Testa, 
Phys. Lett.  {\bf B587}, 157, (2004),  hep-lat/0402027.
\bibitem{Lus2}
M.~L\"uscher, 
Phys. Lett.  {\bf B593}, 296, (2004),  hep-th/0404034.
\bibitem{Edwards:1998wx}
  R.~G.~Edwards, U.~M.~Heller and R.~Narayanan,
  Phys.\ Rev.\  D {\bf 59} (1999) 094510, hep-lat/9811030.
\bibitem{GLWW03} 
L.~Giusti, M.~L\"uscher, P.~Weisz, H.~Wittig, 
JHEP {\bf 11}, 023 (2003), hep-lat/0309189.
\bibitem{DP} 
L.~Del~Debbio, C.~Pica,
JHEP {\bf 0402},  003 (2004), hep-lat/0309145.
\bibitem{DGP04} 
L.~Del~Debbio, L.~Giusti, C.~Pica,
Phys. Rev. Lett. {\bf 94}:032003 (2005), hep-th/0407052.
\bibitem{secondo} 
L.~Giusti, S.~Petrarca, B.~Taglienti,
PoS LAT2006:058,2006,hep-lat/0705.3151.
\bibitem{terzo} 
L.~Giusti, S.~Petrarca, B.~Taglienti,
Phys.Rev. D.76:094510,2007,hep-th/0705.2352.
\bibitem{GHLW02} 
L.~Giusti, C.~Hoelbling, M.~Luscher, H.~Wittig,
Comput. Phys. Commun. {\bf 153}, 31 (2003), hep-lat/0212012.
\bibitem{GRTV}
L.~Giusti, G.~C.~Rossi, M.~Testa, G.~Veneziano,
Nucl. Phys. {\bf B628}, 234, (2002),  hep-lat/0108009.
\end{thebibliography}
\end{document}